# On turbulence:

# deciphering a renormalization flow out of an elliptic curve, II[1]

### LUÍS G.C.D. BORGES[2]

The use of power series to model natural phenomena is a common practice. Sometimes, however, when the practical situation at hand imposes us to acknowledge our own limits, facing infinity, then, one is driven to approximations. This is when the convenience to resort to a polynomial model may appear as a natural, pragmatic way to go ahead.

This is also the reason why, around 1997, trying to reach for a deeper understanding of turbulence, and having thus become interested in studying the dynamics of those formal power series associated with some elliptic curves over $\mathbb{Q}$, I turned my attention to certain ninth degree polynomials, which I found to be canonically deducible from them.

The choice of these special curves, selected for their bad, semi-stable reduction over some common place, was motivated by their being isomorphic to rigid analytic varieties. Conjecturally, this will make them good models, with which to show that there are geodesic paths, characteristic of each and every renormalization flow.

Taking these steps, I was following ideas suggested by my readings of John T. Tate's 1974 paper "The Arithmetic of Elliptic Curves" ([Tate 1974]) and also Joseph H. Silverman's 1986 book, with the same title ([Silverman 1986], pp. 110-115).

The goal was to develop models for turbulence phenomena; more precisely, to use the above mentioned polynomials to model the scale parameterized renormalization flow that is typical of that kind of phenomena. For instance, coherent structures being approximately replicated across the scales, like little vortices becoming bigger and bigger, or smaller and smaller, and so on and so forth.

---

[2] Author's contact e-mail: luiscborges@netcabo.pt .

However, as I proceeded to probe the dynamics of these special polynomials (see fig. 1), I was quite uncertain about what the right choice of an observable would be, with which to put up a classification, discriminating them in a suitable way.

---

$$P(x) = x^3(1 + A_1 x + A_2 x^2 + \cdots + A_6 x^6)$$

(each $A_n$ ($n = 1, 2, \ldots, 6$) denotes a polynomial in the coefficients of the Weierstrass model for a semi-stable elliptic curve over $\mathbb{Q}$)

Fig. 1

---

In fact, since the year 2004, when the first computer simulations were made, no definitive breakthrough was reached, along this path, in spite the observation of some numerical evidence motivating the search.

Meanwhile, concurrent with this search, another line of investigation was to be developed, which is to explore the idea that at least some of the dynamical properties of these polynomials may be encoded in the L functions that come, so to speak, with the former, original elliptic curves, from which they were deduced. The working hypothesis was then put forward that these L functions encode or, more strongly, determine the dynamics of the polynomials with which I proposed myself to model turbulence.

As some experts have pointed out, there is deep arithmetic information encoded in the behavior of these L functions, beyond their region of convergence (a notice I took from Anthony W. Knapp's 1992 book, *Elliptic Curves*, ([Knapp 1992], p. 386). So, my working hypothesis took the form that, in what concerns the relevant dynamics, the same goes, and is to be learned from the region of divergence that is typical of these functions.

In order to test this, I decided myself to look at the way that the many iterates generated with each statistical unit from a sample of thirty L functions explode to infinity. The window of observation was defined as a small strip of the complex plane, including a piece of the border separating convergent behavior from divergent behavior. The first sample was taken among the population of all semi-stable elliptic curves over $\mathbb{Q}$, having bad reduction at the same place ($p = 3$), and with conductor number chosen in a mixed, systematic random way, between 11 and 1000.

In other words, taken from Robert L. Devaney's 2006 work on Complex Exponential Dynamics (published in [Broer et al., eds. 2010]), I turned my attention to some interesting bifurcation diagrams, with my curiosity being partially aroused by the images thereby exposed (see fig. 2).

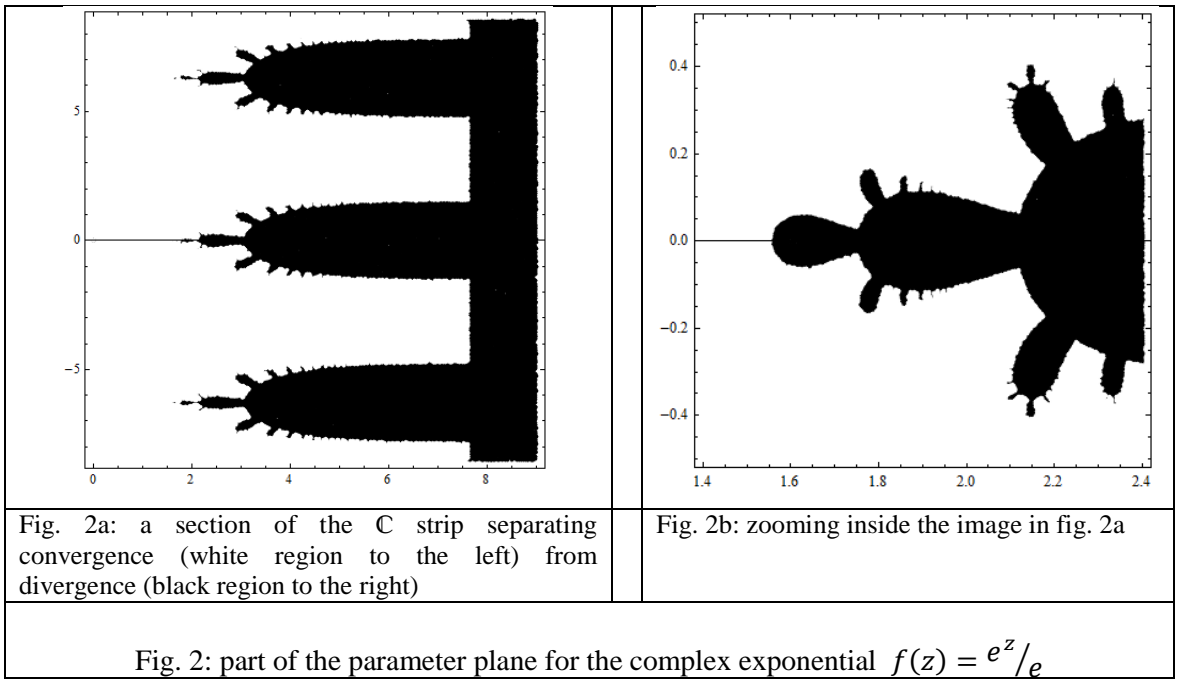

| Fig. 2a: a section of the ℂ strip separating convergence (white region to the left) from divergence (black region to the right) | Fig. 2b: zooming inside the image in fig. 2a |
|---|---|
| Fig. 2: part of the parameter plane for the complex exponential $f(z) = e^z/e$ | |

The plan was to apply the same technique, namely, an escape times algorithm, to render images detailed enough to probe for the existence of some analogous structuring, in the bifurcation diagram for an L function.

Using one the algorithms presented by J. E. Cremona, in his 1997 2nd edition of "Algorithms for Modular Elliptic Curves" [Cremona 1997], and the application for computations Mathematica® 9, from Wolfram Research, I wrote the codes needed to generate the first thousand coefficients in the series representation of each of these thirty, mentioned L functions.

Then, as the correspondent dynamics started to be simulated, by the end of 2008, the following sequence of images of bifurcation diagrams being unveiled started to be documented (see Figs. 3a-3d). They were all obtained under the same numerical conditions and they made it obvious that I had come across something worth of a more systematic exploration then the rendering of interesting pictures.

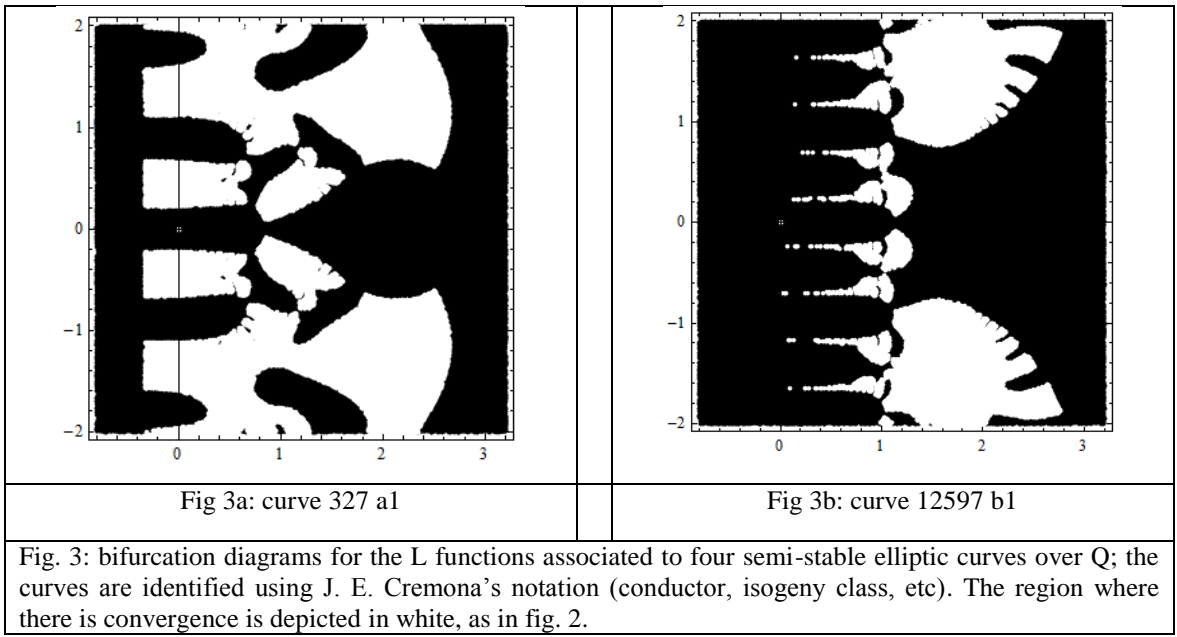

| Fig 3a: curve 327 a1 | Fig 3b: curve 12597 b1 |

Fig. 3: bifurcation diagrams for the L functions associated to four semi-stable elliptic curves over Q; the curves are identified using J. E. Cremona's notation (conductor, isogeny class, etc). The region where there is convergence is depicted in white, as in fig. 2.

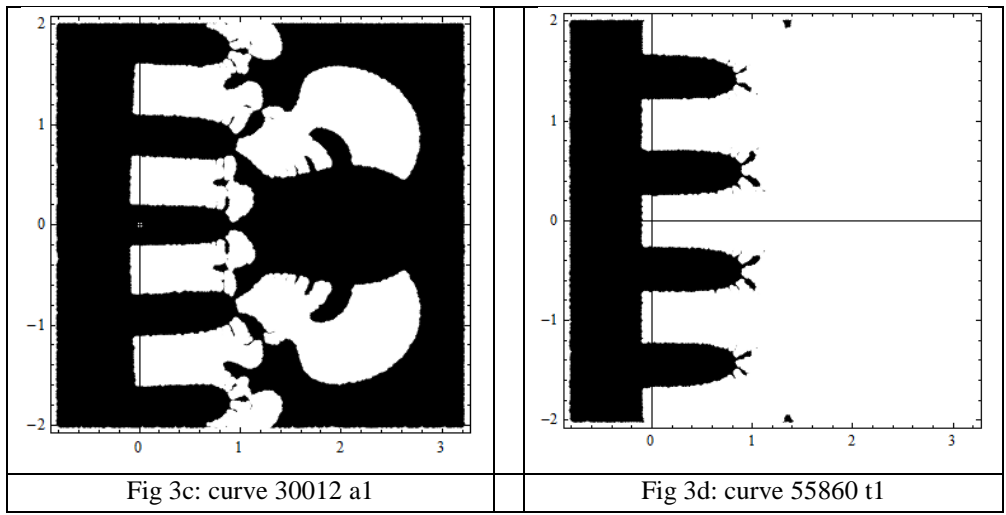

| Fig 3c: curve 30012 a1 | Fig 3d: curve 55860 t1 |

Now, before going further in this narrative, a few inferences are in order and may justify immediate acceptance, as a more close look into what happens with curve 21a1 (see next, Fig. 4), together with the previous images (in Fig. 3), and the correspondent ones for Riemann's zeta function (in Fig. 5), seems to follow by inspection only.

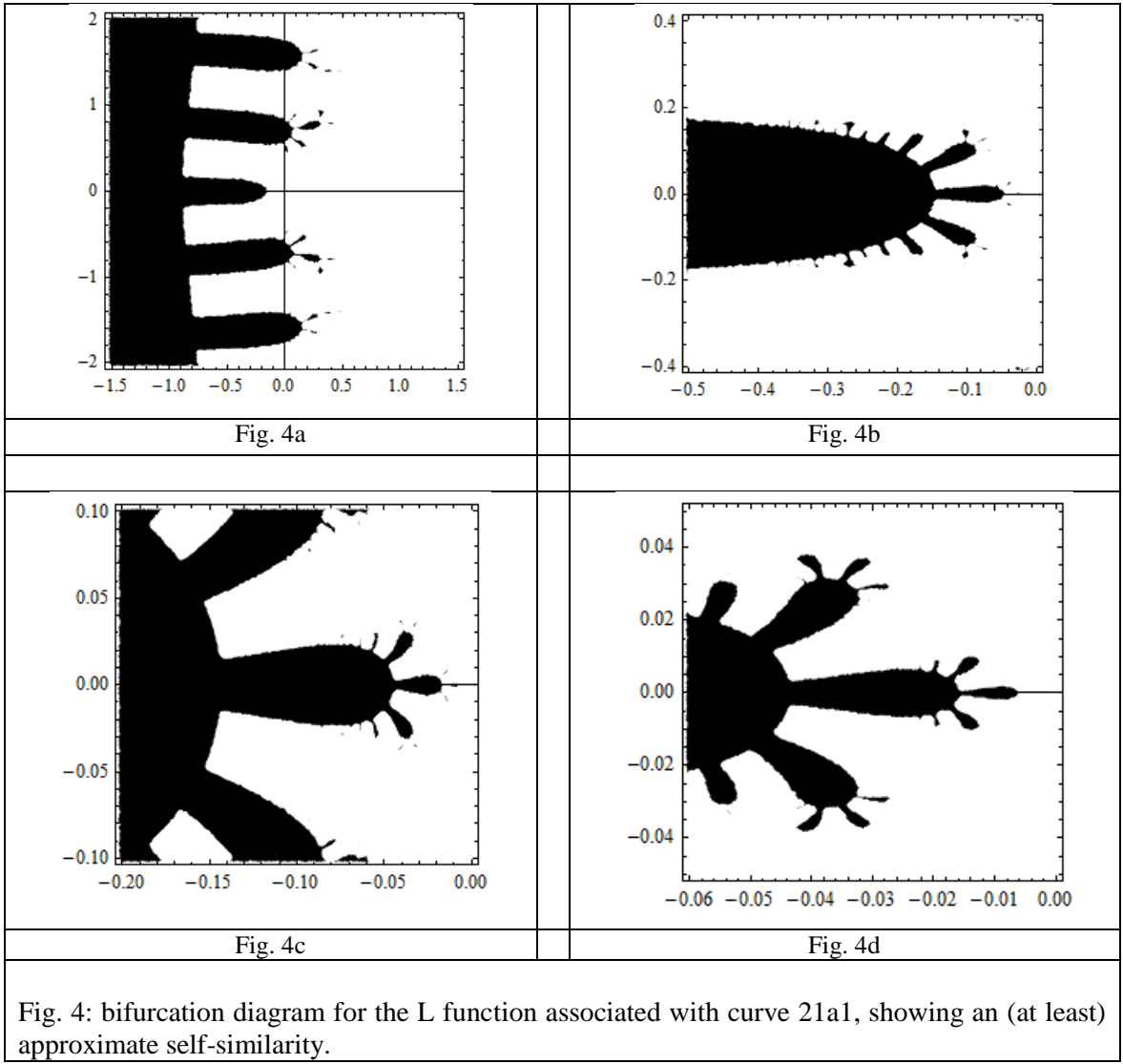

Fig. 4a

Fig. 4b

Fig. 4c

Fig. 4d

Fig. 4: bifurcation diagram for the L function associated with curve 21a1, showing an (at least) approximate self-similarity.

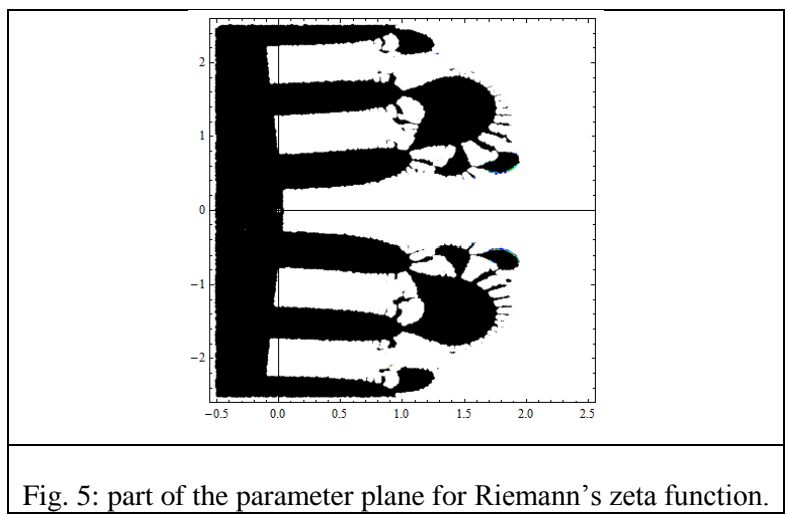

Fig. 5: part of the parameter plane for Riemann's zeta function.

Summing them up, it is clear that:

a) We no longer necessarily have a vertical strip separating a region of convergence from one of divergence, as it is seen that it happens with the complex exponential (Fig. 2) and Riemann's zeta function (Fig. 5); we can have divergence on both sides of the strip (Figs 3a, 3b and 3c);

b) On the other hand, similarly to what happens in the complex exponential case, each extreme of at least some of the black protuberances shown (the 'fingers', in R. L. Devaney's terms), making the interface between convergence and divergence, have in its own boundary an image of the whole strip ( the boundary of the central, main protuberance shown in Fig. 4b reproduces the image shown in Fig. 4a, apart from the main curvature);

c) The particular aspects that are unveiled in the approximate self-similarity shown, from curve to curve, or L function to L function, depend on the coefficients taken into consideration in the numerical approximations that were made to these later functions: the more coefficients one takes into account, the more structure will appear in the bifurcation diagrams, beyond some correspondent or convenient scale threshold.

This last inference can be translated into the following, more sophisticated assertion: the never ending flow of structure, or renormalization flow that one finds, scanning these bifurcation diagrams across the scales, is encoded in the algorithms that model the action of some certain Hecke operators on the Hilbert space spanned by the modular newforms that share these L functions with the curves before-mentioned. This is because the Fourier coefficients of the later can be deduced from the eigenvalues of the former.

By the end of 2012, these inferences were repeatedly validated, as it was possible to extend the numerical evidence supporting them. In fact, as more and more samples were studied, their behavior kept on showing the same, initially observed, rich and diverse phenomena.

Now, back to the point, the question was, and it still is, how to deduce, from these L functions, whatever may help understanding the behavior of their associated ninth degree polynomials. Given that this behavior is assumed to model a few dynamical aspects of turbulence, namely, renormalization flows, it follows as quite a natural option to start with the dynamical behavior of these L functions. I mean, going beyond the image rendering of their bifurcation diagrams and making its dynamics the object of some convenient quantification.

This was the motive why, starting by the end of 2008 and up to the end of 2010, a long trial and error process was to be developed, that took me to try with the escape rate, $\tau$, as an observable. The escape rate is a statistical quantification of how fast the trajectories of a map, in this case, the iterates of an L function explode to infinity ([Tél 1987] and [Beck and Schlögl 1993]).

More specifically, the escape rate, $\tau$, is related to a special value of the topological pressure; in particular, a value that may be interpreted as a free energy for the map under consideration. Thus, the conjecture started to take form, that a variation in this free energy, measured along the above mentioned sample of 30 L functions, might be related in a systematic way to a concomitant variation of some adequate observable, within the set of the respectively, correspondent ninth degree polynomials.

As this research strategy clearly begs for some validation criteria, I decided not to adventure further, at least not before granting myself this approach to be consistent with some canon, within elliptic curve theory. This was the reason why I decided to run the codes previously written for the L functions, in order to make numerical approximations to the values taken by them on the point $z = 1$, in the complex plane. The point $z = 1$, clearly located on the narrow critical strip separating convergence from divergence, is quite a remarkable point on which to study the behavior of L functions, as suggested by the Birch and Swinnerton-Dyer conjecture.

By the end of December 2010, the two lists of values thereby obtained, along the same sample (one list for $L(1)$, the other one for $\tau$) were studied, in search of a statistical connection: their rank correlation coefficient (Spearman coefficient, $r_S$) was determined and found to be $r_S = -0.76$. Afterwards, this result was tested for statistical significance.

Considering the initial sample of size 30, with 28 degrees of freedom, on a level of confidence $\alpha = 0.001$, the null hypothesis that there is no rank correlation between the two lists was rejected (the critical point for this test being $T_{cr} = 0.45$). This kind of correlation was chosen in order to assume the minimum, about $L(1)$ and $\tau$ as statistical variables, when testing for significance.

This result seemed encouraging enough and I proceeded to study a second, larger sample. The second sample was also chosen as in a mixed, systematic random way, among the population of all semi-stable elliptic curves over $\mathbb{Q}$ with bad reduction for $p = 3$, but with conductors now up to 10 000 (taken from [Cremona 2013]).

Just as with the first sample, the same connection, or statistical relation was found, between the escape rates of the associated L-functions and their correspondent values on the point $z = 1$, in the complex plane. Again, this statistical relation took the form of a significant, strong and negative rank correlation, as measured by a Spearman coefficient value $r_s = -0.76$, on a level of confidence $\alpha = 0.001$. The size of this second sample was 70 and I found it worth to go a little deeper and to determine the $p$-value for the test. The result was $p = 3,107 \times 10^{-14}$, making it obvious that I should definitely go beyond the alfa values obtained from common tables.

Thus, from March to May, 2013, using the same data base as source ([Cremona 2013]) a new sample was chosen, with size 325, from the set of all semi-stable elliptic curves with bad reduction at $p = 3$ and conductor number up to 60 000.

The conductor numbers were pseudo-randomly chosen. Whenever there was more than one isogeny class, the same was done to determine from what isogeny class to pick a curve. The curve that was chosen was always the optimal one.

As before, the value of each L function, at $z = 1$, in the complex plane, was determined using truncated series expansions with 1000 coefficients.

The escape rates were all measured in a window of the complex plane defined by $[-1.5, 4.5]$, on the real axis, and $[0, 12]$ on the imaginary axis. Starting from a randomly generated seed inside this window, the criteria for divergence to obtain was based on the crossing over the absolute value 100 000, after 10 iterates. For each escape rate thus determined, exactly 25 000 randomly generated seeds were used.

The statistical association finally found, between the escape rates of the L-functions involved and their respective values on the point $z = 1$, in the complex plane was similar to the previously found ones:

Just as before, a significant, strong and negative, rank correlation was determined, as measured by a Spearman coefficient value $r_S = -0.78$. The $p$-value for this last test was found to be $p = 7{,}9098 \times 10^{-69}$.

Considering the obtained $p$-value, it is natural to reject the null hypothesis that there is no significant statistical association between L function's escape rates and their values at 1. However, the precise nature of the relation leaving such a signature on the statistical ground remains unsettled. The fact that similar values for the rank correlation coefficient were obtained in the preliminary studies, targeting smaller samples, may suggest that the obvious numerical limitations affecting the experimental protocol that was followed are not limitations of a great relevance. In what concerns the quantification of strength, or intensity of the correlation that was measured, an unknown factor may well exist, that systematically prevented obtaining a rank correlation coefficient close to 1, apart from numerical limitations.

This work illustrates the relevance that the use of computers may have, in collecting information that even a simple statistical analysis may convert into the knowledge of a non-trivial connection, between different areas of Mathematics. From a numerical and an experimental perspective, it validates the research strategy that was followed, electing the escape rate as an observable with which to discriminate between L functions.

Lisbon, 24/7/2013